\RequirePackage{fix-cm}
\documentclass[smallextended]{svjour3}       % onecolumn (second format)
\smartqed  % flush right qed marks, e.g. at end of proof
\usepackage{graphicx}
\begin{document}

\title{Decay properties of singly charmed baryons %Strong and radiative decays of singly charmed baryons%\thanks{Grants or other notes
}
%\address{}
%\subtitle{Department of Applied Physics, Sardar Vallabhbhai National Institute of Technology, Surat 395007, Gujarat, India.}


\author{Keval Gandhi{$^a$}, Zalak Shah{$^b$} and Ajay Kumar Rai{$^c$}     
         %etc.
}

\authorrunning{K. Gandhi, Z. Shah and A. K. Rai} % if too long for running head

\institute{\at
             Department of Applied Physics, Sardar Vallabhbhai National Institute of Technology, Surat$-$395007, Gujarat, India. \\
              $^a$\email{keval.physics@yahoo.com}           %  \\
           \at 
              %Department of Applied Physics, Sardar Vallabhbhai National Institute of Technology, Surat$-$395007, Gujarat, India.\\
              %second address\\
             $^b$\email{zalak.physics@gmail.com}           %  \\
            \at
               %Department of Applied Physics, Sardar Vallabhbhai National Institute of Technology, Surat$-$395007, Gujarat, India.\\
              %second address\\
             $^c$\email{raiajayk@gmail.com}           %  \\
}

\date{Received: date / Accepted: date}

\maketitle

\begin{abstract}
The magnetic moments, transition magnetic moments and the radiative decay widths of singly charmed baryons are  calculated  with $ {J^P = \frac{1}{2}}^+$ and $ {J^P = \frac{3}{2}}^+$ in the constitute quark model. Further, the strong decay rates for $ S $, $ P $ and $ D $ wave transitions are also presented. The singly charmed baryon masses used in the calculations were obtained from the hypercentral Constitute Quark Model (hCQM) without and with first order relativistic correction. Obtained results are compared with experimental observation as well as with the other theoretical predictions.
\end{abstract}

\section{Introduction}
\label{intro}

The ground state masses of singly charmed baryons are well established and many of their radially and orbitally excited states masses are well-known experimentally \cite{partignani2016c} as well as theoretically in our previous work \cite{shah2016excited}. In order to understand the structural properties of the singly charmed baryons, it is necessary to analyze the decay modes from theoretical study. An experimental observations for the radiative decay of singly charmed baryons are rare; whereas their strong decay rates, widths and lifetimes are measured by various experimental groups \cite{lee2014measurements,aaltonen2011measurements,athar2005new,artuso2002measurement,link2000measurements,crawford1993observation,yelton2016study,avery1995observation,albrecht1997evidence,edwards1995observation} till the date. The various properties of heavy baryons are nicely presented in these review articles \cite{cheng2015charmed,chen2017review,crede2013progress,klempt2010baryon}.

In order to improve the structural understanding of baryons (made of both light and heavy quarks) 
the magnetic moment is an important tool. There are many theoretical approaches which study the individual contribution of quarks in the magnetic moments of baryons; such as, heavy chiral perturbation theory \cite{li2017magnetic,Wang:2018gpl}, effective quark mass scheme \cite{dhir2009magnetic}, bag model \cite{bernotas2013radiative}, QCD sum rule model \cite{aliev2009mass}, lattice QCD \cite{can2014electromagnetic,bahtiyar2015omegacgamma,bahtiyar2017xicgamma}, relativistic quark model \cite{faessler2006magnetic,barik1983magnetic}, non-relativistic quark model \cite{albertus2007static,0954-3899-35-6-065001}, chiral constitute quark model \cite{sharma2010spin} etc. For the radiative decay, there is no phase space and isospin conservation constraint for the transitions of mass-less photon among the charmed baryons. There are many phenomenological approaches; relativistic quark model \cite{ivanov1999strong}, bag model \cite{bernotas2013radiative}, QCD sum rule model \cite{aliev2009radiative,aliev2012vector}, non-relativistic constitute quark model \cite{shah2016mass,majethiya2009radiative,wang2017strong}, heavy hadron chiral perturbation theory \cite{jiang2015electromagnetic,cheng1997remarks,cheng1993chiral,cho1994strong} etc. have  calculated the contribution of radiative interaction in the decay of singly charmed baryons. The future experiments at J-PARC, $ \bar{P}ANDA $ \cite{B. Singh(2017),B. Singh(2016a),B. Singh(2016),B. Singh(2017a),lutz2016resonances} and LHCb are expected to give further information on charmed baryons.

The fundamental theory of the strong interactions, Quantum Chromodynamics (QCD), simplifies enormously in the presence of a system containing one heavy quark $(c$ or $b)$ and two light quarks $(u$, $d$ or $s)$. It will provide the understanding of the SU(4) spin-flavor symmetry of heavy quark and the SU(3) symmetry of light quarks. Such a heavy quark symmetry arises when the mass of the heavy quark is much larger than the QCD limit $ \Lambda_{QCD} \simeq 0.2 $ GeV \cite{neubert1994heavy}. In this heavy quark limit the dynamics of heavy and light quarks are decouple and providing a number of model independent relations between various decay mode of the heavy baryons. The chiral Lagrangian corresponding to the heavy baryon coupling to the pseudoscalar mesons were first introduced in Ref.\cite{yan1992heavy} in 1992. Theoretically, the relativistic constitute quark model \cite{ivanov1999strong}, the non-relativistic quark model with various QCD inspired potentials \cite{albertus2005study,huang1995decays,cheng2015charmed}, light-front quark model \cite{tawfiq1998charmed}, Heavy Hadron Chiral Perturbation Theory (HHCPT) \cite{cho1994strong,yan1992heavy,pirjol1997predictions}, and the QCD sum rules on the light cone \cite{agaev2017nature} etc. are used for studying the strong decays of singly charmed baryons by an exchange of a single pion. 

This paper is organized as follows: The basic methodology adopted for generating the mass spectra of singly charmed baryons is described in section II. The magnetic moments and the electromagnetic radiative decays from their transition magnetic moments of ground state with  $ J^P=\frac{1}{2}^+ $ and $ J^P=\frac{3}{2}^+ $ are presented in section III.  The details of hadronic strong decays of singly charmed baryon are presented in section IV. In the last section, we draw our discussion and conclusion.
\begin{table}
\caption{\label{tab:1}The masses of singly charm baryons \cite{shah2016excited} (in MeV).}
\label{tab:1}       % Give a unique label
\begin{tabular}{cccccccc}
\hline\noalign{\smallskip}
Baryon & State & $ M_A $ & $ M_B $ & PDG \cite{partignani2016c} \\
 &  $n$ $^{2S+1}L_{J} $ & & & & \\
\hline

$\Lambda{_{c}^{+}}$ & 1$^{2}S_{\frac{1}{2}} $ & 2286 & 2286 & 2286.46 $ \pm $ 0.14 \\

$\Sigma{_{c}^{++}}$ & 1$^{2}S_{\frac{1}{2}} $ & 2449 & 2454 & 2453.97 $ \pm $ 0.14 \\

$\Sigma{_{c}^{+}}$ & 1$^{2}S_{\frac{1}{2}} $ & 2444 & 2452 & 2452.9 $ \pm $ 0.4 \\

$\Sigma{_{c}^{0}}$ & 1$^{2}S_{\frac{1}{2}} $ & 2444 & 2453 & 2453.75 $ \pm $ 0.14 \\

$\Xi{_{c}^{+}}$ & 1$^{2}S_{\frac{1}{2}} $ & 2467 & 2467 & 2467.87 $ \pm $ 0.30 \\

$\Xi{_{c}^{0}}$ & 1$^{2}S_{\frac{1}{2}} $ & 2470 & 2470 & 2470.87 $ {^{+0.28}_{-0.31}} $ \\

$\Omega{_{c}^{0}}$ & 1$^{2}S_{\frac{1}{2}} $ & 2695 & 2695 & 2695.2 $ \pm $ 1.7 \\

\hline\noalign{\smallskip}

$\Sigma{_{c}^{*++}}$ & 1$^{4}S_{\frac{3}{2}} $ & 2505 & 2530 & 2518.41 $ {^{+0.21}_{-0.19}} $ \\

$\Sigma{_{c}^{*+}}$ & 1$^{4}S_{\frac{3}{2}} $ & 2506 & 2501 & 2517.5 $ \pm $ 2.3 \\

$\Sigma{_{c}^{*0}}$ & 1$^{4}S_{\frac{3}{2}} $ & 2506 & 2529 & 2518.48 $ \pm $ 0.20 \\

$\Xi{_{c}^{*+}}$ & 1$^{4}S_{\frac{3}{2}} $ & 2625 & 2619 & 2645.53 $ \pm $ 0.31 \\

$\Xi{_{c}^{*0}}$ & 1$^{4}S_{\frac{3}{2}} $ & 2584 & 2610 & 2646.32$ \pm $ 0.31 \\

$\Omega{_{c}^{*0}}$ & 1$^{4}S_{\frac{3}{2}} $ & 2740 & 2745 & 2765.9 $ \pm $ 2.0 \\

\hline

$\Lambda{_{c}^{+}}$ & 1$^{2}P_{\frac{1}{2}} $ & 2607 & 2692 & 2592.25 $ \pm $ 0.28 \\

$\Sigma{_{c}^{++}}$ & 1$^{2}P_{\frac{1}{2}} $ & 2842 & 2890 & $ 2801{^{+4}_{-6}}$  \\

$\Sigma{_{c}^{+}}$ & 1$^{2}P_{\frac{1}{2}} $ & 2831 & 2849 & $ 2792{^{+14}_{-5}}$  \\

$\Sigma{_{c}^{0}}$ & 1$^{2}P_{\frac{1}{2}} $ & 2824 & 2873 & 2806 $ {^{+5}_{-7}}$ \\

\hline\noalign{\smallskip}

$\Lambda{_{c}^{+}}$ & 1$^{2}P_{\frac{3}{2}} $ & 2592 & 2612 & 2628.11 $ \pm $ 0.19 \\

$\Sigma{_{c}^{++}}$ & 1$^{2}P_{\frac{3}{2}} $ & 2814 & 2860 & $ - $ \\ 

\hline\noalign{\smallskip}

$\Sigma{_{c}^{++}}$ & 1$^{4}P_{\frac{5}{2}} $ & 2791 & 2835 & $ - $ \\ 

\noalign{\smallskip}\hline
\footnote{z}
\end{tabular}

$ M_A $ $ \rightarrow$ without first order correction masses

$ M_B $ $ \rightarrow$ with first order correction masses.
\end{table}
\section{Methodology}

The mass spectra of singly charmed baryons \cite{shah2016excited,zalakepjc(2016),zalakepjc(2017),zalakepja(2017)} are generated by the Hamiltonian
\begin{equation}
H =\frac{P{^{2}_x}}{2m} + V(x)
\end{equation}
\noindent in the hypercentral Constitute Quark Model (hCQM). Here, $ m = \frac{m_\rho m_\lambda}{m_\rho + m_\lambda}$ is the reduced mass and $x$ is the six dimensional radial hypercentral coordinate of the three body system. In this case, we consider the hypercentral potential $ V(x) $ as the color Coulomb plus power potential with first-order correction as
\begin{equation}
V(x) = V^0(x) + \left(\frac{1}{m_\rho} + \frac{1}{m_\lambda}\right) V^1(x) + V_{SD}(x)
\end{equation}
where $ V_{SD}(x) $ represents the spin dependent potential, $V^0(x)$ is the sum of hyper Coulumb (hC) interaction and a confinement term
\begin{equation}
V^0(x) = \frac{\tau}{x} + \beta x
\end{equation}
\noindent and the first order correction is employed by Koma et al \cite{Koma(2006)};
\begin{equation}
V^1(x) = -C_F C_A \frac{\alpha_{s}^2}{4x^2}.
\end{equation}
\noindent We have used this correction not only for baryons but mesons as well \cite{V.Kher(2018)}. Here, the hyper-Coulumb strength $ \tau = - \frac{2}{3} \alpha_s $; where $\frac{2}{3}$ is the baryon color factor and $ \alpha_s $ represents the strong running coupling constant and is $\approx$ 0.6 considered in the present study. $ \beta $ is the string tension of the confinement; and $ C_F $ and $ C_A $ are the Casimir charges of the fundamental and adjoint representation. The details of all the constants can be found from Ref.\cite{shah2016excited}.

For the quarks $u, d, s$ and $ c $; we set the constituent quark masses $m_u=338$MeV, $m_d=350$MeV, $m_s=500$MeV and $m_c=1275$MeV. The 1S and 1P state masses of singly charmed baryons are tabulated in Table \ref{tab:1} with PDG masses \cite{partignani2016c}. $ M_A $ and $ M_B $ are the masses of without and with first order relativistic correction to the potential energy term, respectively. We will use these masses in the calculation of magnetic moments, the radiative decays and the strong decays in next sections.

\section{Magnetic Moments and Radiative Decays}
The magnetic moments and the radiative decays are computed using spin-flavour wave functions of the participating baryons. The magnetic moments are obtained in terms of spin, charge and effective mass of the bound quarks of baryons. In radiative decay, there is an exchange of massless photon among the singly charmed baryons. Such a decay does not contained phase space restriction. Therefore, some of the radiative decay mode of heavy baryons are contributed significantly to the total decay rate.

\subsection{The Magnetic Moments}
\begin{table*}
\begin{center}
\caption{\label{tab:2}Magnetic moments of the singly charmed baryons with  $ J^P = {\frac{1}{2}}^+ $ (in $ \mu_N $).}
\scalebox{0.75}{
\begin{tabular}{ccccccccccccccccccc}
\hline\noalign{\smallskip}
Baryon & Expression & A & B & \cite{Wang:2018gpl}  & \cite{can2014electromagnetic,bahtiyar2015omegacgamma,bahtiyar2017xicgamma} & \cite{bernotas2013radiative} & \cite{dhir2009magnetic} & \cite{patel2008heavy} & \cite{sharma2010spin} & \cite{faessler2006magnetic}  \\
\hline\noalign{\smallskip}
$ \Lambda{_{c}^{+}} $ & $\mu_c$ & 0.421 & 0.421 & 0.21 & & 0.411 & 0.370 & 0.385 & 0.39 & 0.42 \\

$ \Sigma{_{c}^{++}} $ &$\frac{4}{3}\mu_u - \frac{1}{3}\mu_c$ & 1.835 & 1.831 & 1.50 & $ 1.499(202) $ & 1.679 & 2.09 & 2.279 & 2.540 & 1.76 \\

$ \Sigma{_{c}^{0}} $ & $\frac{4}{3}\mu_d - \frac{1}{3}\mu_c$ & -1.095 & -1.091 & $ - $1.25 & $ -0.875(103) $ & -1.043 & -1.230 & -1.015 & -1.46 & -1.04 \\

$ \Sigma{_{c}^{+}} $ & $\frac{2}{3}\mu_u + \frac{2}{3}\mu_d - \frac{1}{3}\mu_c$ & 0.381 & 0.380 & 0.12 &  & 0.318 & 0.550 & 0.500 & 0.540 & 0.36 \\

$ \Xi{_{c}^{0}} $ & $\frac{2}{3}\mu_d + \frac{2}{3}\mu_s - \frac{1}{3}\mu_c$ & -1.012 & -1.012 & 0.19 & $ 0.192(17) $ & -0.914 & -0.940 & -0.966 & -1.23 & \\

$ \Xi{_{c}^{+}} $ & $\frac{2}{3}\mu_u + \frac{2}{3}\mu_s - \frac{1}{3}\mu_c$ & 0.523 & 0.523 & 0.24 & $ 0.235(25) $ & 0.591 & 0.75 & 0.711 & 0.770 & 0.41 \\

$ \Omega{_{c}^{0}} $ & $\frac{4}{3}\mu_s - \frac{1}{3}\mu_c$ & -1.127 & -1.179 & $ - $ 0.67 & $ -0.667(96) $ & -0.774 & -0.890 & -0.960 & -0.900 & -0.85 \\

\hline\noalign{\smallskip}
\end{tabular}
}
\end{center}

\end{table*}

\begin{table*}
\begin{center}
\caption{\label{tab:3}Magnetic moments of the singly charmed baryons with $ J^P = {\frac{3}{2}}^+ $ (in $ \mu_N $).}
\scalebox{0.85}{
\begin{tabular}{ccccccccccccccccccc}
\hline\noalign{\smallskip}
Baryon & Expression & A & B  & \cite{bernotas2013radiative} & \cite{dhir2009magnetic} & \cite{patel2008heavy} & \cite{sharma2010spin} & \cite{albertus2007static} & \cite{aliev2009mass} \\
\hline\noalign{\smallskip}

$ \Sigma{_{c}^{*++}} $ & $ 2 \mu_u + \mu_c $ & 3.264 & 3.232  & 3.127 & 3.630 & 3.844 & 4.390  & & $4.81 \pm 1.22 $\\

$ \Sigma{_{c}^{*+}} $ & $ \mu_u + \mu_d + \mu_c $ & 1.134 & 1.136 & 1.085 & 1.180 & 1.256 & 1.390 && $ 2.00 \pm 0.46 $\\

$ \Sigma{_{c}^{*0}} $ & $ 2 \mu_d + \mu_c $ & -1.054 & -1.044  & -0.958 & -1.180 & -0.850 & -1.610  & -1.99 & $ - 0.81 \pm 0.20 $\\

$ \Xi{_{c}^{*0}} $ & $ \mu_d + \mu_s + \mu_c $ & -0.846 & -0.837  & -0.746 & -1.020 & -0.690 & -1.260  & -1.49\\

$ \Xi{_{c}^{*+}} $ & $ \mu_u + \mu_s + \mu_c $ & 1.330 & 1.333  & 1.270 & -1.390 & 1.517 & 1.740 &  & $ 1.68 \pm 0.24 $\\

$ \Omega{_{c}^{*0}} $ & $ 2 \mu_s + \mu_c $ & -1.127 & -1.129  & -0.547 & -0.840 & -0.867 & -0.910  & -0.860 & $ - 0.62 \pm 0.18 $\\

\hline\noalign{\smallskip}

\end{tabular}
}
\end{center}

\end{table*}

The magnetic moment is the fundamental property of baryon in both light and heavy quark sector and is purely depends upon the masses and spin of their internal constitutions. The magnetic moment of the baryon $ (\mu_B) $ is given by the expectation value \cite{0954-3899-35-6-065001,shah2016mass} as 

\begin{equation}
\mu_B = \sum_{q} \left\langle \Phi_{sf} \middle| \mu{_q}_z \middle| \Phi_{sf} \right\rangle  ;q=u,d,s,c.
\end{equation}  

\noindent where, $ \Phi_{sf} $ represents the spin-flavour wave function of a participating baryon and $ \mu{_q} $ is the magnetic moment of the individual quark given by 

\begin{equation}
\mu_{q} = \frac{e_q}{2 m{_{q}^{eff}}} \cdot \sigma_q
\end{equation}  

\noindent with $ e_q $ is the charge and $ \sigma_q $ is the spin of the constitute quark of the particular baryonic state, and the effective mass of each constituting quark $ (m{_{q}^{eff}}) $ can be defined in terms of the constituting quark mass $ (m{_{q}}) $ as

\begin{equation}
m{_{q}^{eff}} = m{_{q}} \left({1 + \frac{\left\langle H \right\rangle}{\sum\limits_{q} m{_{q}}}}\right)  
\end{equation}

\noindent where, the Hamiltonian is given in the form of measured or predicted baryon mass $ (M) $ as, $ {\left\langle  H \right\rangle} = M - {\sum\limits_{q} m{_{q}}} $. Here, the $ m{_{q}^{eff}} $  represents the mass of the bound quark inside the baryons by taking into account its binding interactions with other two quarks described in Eqn.(1) in the case of hCQM.\\

\noindent Using these equations and taking the constituent quark mass of \cite{shah2016excited}, we determine the ground state magnetic moment of the singly charmed baryons with $ J^P = \frac{1}{2}^+ $ and $ J^P = \frac{3}{2}^+ $ by considering without and with first order relativistic correction as a set A and set B respectively. We present our results in Table \ref{tab:2}-\ref{tab:3} in the unit of nuclear magnetons $\left(\mu_N = \frac{e\hbar}{2m_p}\right)$.

\subsection{The Radiative Decays}

\begin{table*}
\begin{center}
\caption{\label{tab:4}The transition magnetic moments $ \left\vert \mu{_{{B_c} \rightarrow B{_{c}^{\prime}}}} \right\vert $ of singly charmed baryons (in $\mu_N $).}
\begin{tabular}{ccccccccccccccccccc}

\hline\noalign{\smallskip}

Transition & Expression & A & B & \multicolumn{3}{c}{\cite{dhir2009magnetic} } & \cite{majethiya2009radiative} & \cite{aliev2009radiative}  \\
&&& & (nqm) & (ems) & (ses)  &  & & \\

\hline\noalign{\smallskip}
$ \mu_{{\Sigma{_{c}^{+}} \rightarrow \Lambda{_{c}^{+}}}} $ & $ \frac{-1}{\sqrt{3}} (\mu_u - \mu_d) $  & 1.2722 & 1.2680 & 2.28 & 2.28 & 2.15 & 1.347 & 1.48 $\pm$ 0.55 \\

$ \mu_{{{\Sigma{_{c}^{*++}}} \rightarrow \Sigma{_{c}^{++}}}} $ & $ \frac{2\sqrt{2}}{3} (\mu_u - \mu_c) $ & 0.9984 & 0.9885 & 1.41 & 1.19 & 1.23 & 1.080 &1.06 $ \pm $ 0.38 \\

$ \mu_{{\Sigma{_{c}^{*+}} \rightarrow \Sigma{_{c}^{+}}}} $ & $ \frac{\sqrt{2}}{3} (\mu_u + \mu_d - 2\mu_c) $ & 0.0089 & 0.0089 & 0.09 & 0.04 & 0.08 & 0.008 & 0.45 $ \pm $ 0.11\\

$ \mu_{{\Sigma{_{c}^{*0}} \rightarrow \Sigma{_{c}^{0}}}} $ & $ \frac{2\sqrt{2}}{3} (\mu_d - \mu_c) $ & 1.0220 & 1.0127 & 1.22 & 1.11 & 1.07 & 1.064 & 0.19 $ \pm $ 0.08  \\

$ \mu_{{\Sigma{_{c}^{*+}} \rightarrow \Lambda{_{c}^{+}}}} $ & $ \sqrt{\frac{2}{3}} (\mu_u - \mu_d) $  & 1.7546 & 1.7582 & & & & 1.857 & \\

$ \mu_{{\Xi{_{c}^{*+}} \rightarrow \Xi{_{c}^{+}}}} $ & $ \sqrt{\frac{2}{3}} (\mu_u - \mu_s) $ & 0.9832 & 0.9852 & 2.02 & 1.96 & 1.94 & 0.991 & 1.47 $ \pm $ 0.66 \\

$ \mu_{{\Xi{_{c}^{*0}} \rightarrow \Xi{_{c}^{0}}}} $ & $ \sqrt{\frac{2}{3}} (\mu_d - \mu_s) $ & 0.2552 & 0.2527 & 0.26 & 0.25 & 0.18 & 0.120 & 0.16 $ \pm $ 0.07 \\

$ \mu_{{\Omega{_{c}^{*0}} \rightarrow \Omega{_{c}^{0}}}} $ & $ \frac{2\sqrt{2}}{3} (\mu_s - \mu_c) $ & 0.8734 & 0.8719 & 0.92 & 0.88 & 0.90 & 0.908 & \\
\hline\noalign{\smallskip}
\end{tabular}
\end{center}
\end{table*}

\begin{table*}
\begin{center}
\caption{\label{tab:5}The radiative decay widths ($ \Gamma_\gamma $) of singly charmed baryons (in keV).}
\scalebox{0.85}{
\begin{tabular}{ccccccccccccccccccc}

\hline\noalign{\smallskip}

Decay Mode & A & B & \cite{wang2017strong} & \cite{jiang2015electromagnetic} & \cite{aliev2012vector} & \cite{majethiya2009radiative} & \cite{dey1994radiative} & \cite{bernotas2013radiative} & \cite{cheng1997remarks} & \cite{ivanov1999strong} & \cite{aliev2009radiative} \\

\hline\noalign{\smallskip}

$ \Sigma{_{c}^{+}} $ (1$^{2}S_{\frac{1}{2}}) $ $ \rightarrow \Lambda{_{c}^{+}} \gamma $ & 58.131 & 66.660 & 80.6 & 164 & & 60.55 & 120 & 46.1 & 88 & 60.7 $ \pm $ 1.5\\

$ \Sigma{_{c}^{*++}} $ (1$^{4}S_{\frac{3}{2}}) $ $ \rightarrow \Sigma{_{c}^{++}} \gamma $ & 0.8504 & 2.0597 & 3.94 & 11.6 & 3.567 & 1.15 & 1.6 & 0.826 & 1.4 & & 2.65 $ \pm $ 1.60 \\

$ \Sigma{_{c}^{*+}} $ (1$^{4}S_{\frac{3}{2}}) $  $ \rightarrow \Sigma{_{c}^{+}} \gamma $ & 9 $\times$ 10$^{-5}$ & 4 $\times$ 10$^{-5}$ & 0.004 & 0.85 & 0.187 & 0.00006 & 0.0001 & 0.004 & 0.002 & 0.14 $ \pm $ 0.004 & 0.40 $ \pm $ 0.22\\

$ \Sigma{_{c}^{*0}} $ (1$^{4}S_{\frac{3}{2}}) $ $ \rightarrow \Sigma{_{c}^{0}} \gamma $ & 1.2049 & 2.1615 & 3.43 & 2.92 & 1.049 & 1.12 & 1.2 & 1.08 & 1.2 & & 0.08 $ \pm $ 0.042\\

$ \Sigma{_{c}^{*+}} $ (1$^{4}S_{\frac{3}{2}}) $ $ \rightarrow \Lambda{_{c}^{+}} \gamma $ & 143.97 & 135.30 & 373 & 893 & 409.3 & 154.48 & 310 & 126 & 147 & 151 $ \pm $ 4 & 130 $ \pm $ 65   \\

$ \Xi{_{c}^{*+}} $ (1$^{4}S_{\frac{3}{2}}) $ $ \rightarrow \Xi{_{c}^{+}} \gamma $ & 17.479 & 15.686 & 139 & 502 & 152.4 & 63.32 & 71 & 44.3 & 54 & 54 $ \pm $ 3 & 52 $ \pm $ 32\\

$ \Xi{_{c}^{*0}} $ (1$^{4}S_{\frac{3}{2}}) $ $ \rightarrow \Xi{_{c}^{0}} \gamma $ & 0.4535 & 0.8114 & 0.0 & 0.36 & 1.318 & 0.30 & 1.7 & 0.908 & 1.1 & 0.68 $ \pm $ 0.04 & 0.66 $ \pm $ 0.41 \\

$ \Omega{_{c}^{*0}} $ (1$^{4}S_{\frac{3}{2}}) $ $ \rightarrow \Omega{_{c}^{0}} \gamma $ & 0.3408 & 0.4645 & 0.89 & 4.82 & 1.439 & 2.02 & 0.71 & 1.07 \\

\hline\noalign{\smallskip}

\end{tabular}
}
\end{center}
\end{table*}

The electromagnetic radiative decay width is mainly the function of radiative transition magnetic moment $ \mu_{B_{C}^{\prime}\rightarrow B_{c}} $ (in $ \mu_N $) and photon energy ($ k $) \cite{shah2016mass,majethiya2009radiative,patel2008heavy} as

\begin{equation}
\Gamma_\gamma = \frac{k^3}{4\pi}\frac{2}{2J+1}\frac{e}{m{_{p}^2}}\mu{_{{B{_{c}}} \rightarrow {B{_{c}^{\prime}}}}^2}
\end{equation}

where $ m_p $ is the mass of proton, $ J $ is the total angular momentum of the initial baryon $ ({B{_{c}}}) $. Such a transition magnetic moments $ (\mu_{B_c \rightarrow B{_{c}^{\prime}}}) $ are determine in the same manner by sandwiching Eqn.(7) between the appropriate initial $ (\Phi_{sf_{B{_{c}}}}) $ and final state $ (\Phi_{sf_{B{_{c}^{\prime}}}}) $ singly charm baryon spin-flavour wave functions as

\begin{equation}
\mu{_{{B_c} \rightarrow B{_{c}^{\prime}}}} = \langle\Phi_{sf_{B{_{c}}}} \vert \mu{_B}_{c^{\prime}_{z}} \vert \Phi_{sf_{B{_{c}^{\prime}}}} \rangle
\end{equation}

\noindent To determine the  radiative decay of the channel $  \Sigma{_{c}^{*+}} \rightarrow \Lambda{_{c}^{+}} \gamma $, first to need to calculate the transition magnetic moment given as,
 
\begin{equation}
\mu_{ \Sigma{_{c}^{*+}} \rightarrow \Lambda{_{c}^{+}}} = \left \langle\Phi_{sf_{\Sigma{_{c}^{*+}}}} \right \vert \mu_{\Lambda_{{c}^{+}_{z}}} \left\vert \Phi_{sf_{\Lambda{_{c}^{+}}}} \right\rangle
\end{equation}

\noindent the spin-flavour wave functions $ ( \Phi_{sf} ) $ of $  \Sigma{_{c}^{*+}} $ and $ \Lambda{_{c}^{+}} $ baryons are expressed as

\begin{equation}
\left\vert \Phi_{sf_{\Sigma{_{c}^{*+}}}} \right\rangle  = \left(\frac{1}{\sqrt{2}}(ud+du)c\right) \cdot \left(\frac{1}{\sqrt{3}}(\uparrow\uparrow\downarrow+\uparrow\downarrow\uparrow+\downarrow\uparrow\uparrow)\right)
\end{equation}

\begin{equation}
 \left\vert \Phi_{sf_{\Lambda{_{c}^{+}}}} \right\rangle  = \left(\frac{1}{\sqrt{2}}(ud-du)c\right) \cdot \left(\frac{1}{\sqrt{2}}(\uparrow\downarrow-\downarrow\uparrow)\uparrow\right).
\end{equation}

\noindent  Following the orthogonality condition of quark flavour and the spin states, for example $ \left\langle u\uparrow d \uparrow c\downarrow \vert u \uparrow d\downarrow c\uparrow \right\rangle = 0 $; we will get the expression of transition magnetic moment as

\begin{equation}
\mu_{ \Sigma{_{c}^{*+}} \rightarrow \Lambda{_{c}^{+}}} = \sqrt{\frac{2}{3}} \left (\mu_u - \mu_d \right).  
\end{equation}

\noindent The transition magnetic moments are given in Table \ref{tab:4}.  Using the masses and transition magnetic moment of the participating baryons, we compute its radiative decay width. The obtained results are listed in Table \ref{tab:5} for both set A and set B with other theoretical predictions. 

\section{The Strong Decays}
\label{sec:1}

The effective coupling constant of the heavy baryons are small which leads their strong interactions  perturbatively and easier to understand the systems containing only light quarks. Such a theory describe the strong interactions in the low energy regime by an exchange of light Goldstone boson is developed well by the co-ordination of chiral perturbation theory and heavy quark effective theory called Heavy Hadron Chiral Perturbation Theory (HHCPT). This hybrid effective theory has been applied to study the strong and the electromagnetic decays of ground and excited states in the both charm and bottom sector \cite{wise1992chiral,yan1992heavy,burdman1992union}. By using the Langrangian of Ref.\cite{pirjol1997predictions}, we calculated a strong $P-$ wave couplings among the $s-$wave baryons, $S-$wave couplings between the $s-$wave and $p-$wave baryons, and the strong couplings of $D-$wave from $p-$wave to $s-$wave baryons in this section. Such a chiral Lagrangian gives the expressions of typical decay rate of single pion transitions between singly charmed baryons mentiond in Eqns. (15-20) \cite{cheng2015charmed}. The pion momentum for the two body decay $ x \rightarrow y + \pi $ is 

\begin{equation}
p_{\pi}= \frac{1}{2m_x}\sqrt{[m{_{x}^{2}} - (m_y + m_{\pi})^2][m{_{x}^{2}} - (m_y - m_{\pi})^2]}.
\end{equation}\\

\begin{itemize}

\item P$-$wave transitions \\
\label{sec:2}

The decay rates corresponding to the P$-$wave transitions from the isospin partners of $ \Sigma{_{c}}(1^2S{_{\frac{1}{2}}}) $ and  $ \Sigma{_{c}^{*}}(1^{4}S_{\frac{3}{2}}) $ to the state $ \Lambda{_{c}^{+}} (1^{2}S_{\frac{1}{2}}) $ by an exchange of single pion are 

\begin{equation}
\Gamma_{\Sigma{_{c}^+}/\Sigma{_{c}^{*}} \rightarrow \Lambda{_{c}^{+}}\pi} = \frac{a{_{1}^{2}}}{2\pi f{_{\pi}^{2}}} \frac{M_{\Lambda{_{c}^{+}}}}{M_{\Sigma{_{c}^{+}}/\Sigma{_{c}^{*}}}} p{_{\pi}^{3}}
\end{equation}

where $ p{_{\pi}^3} $ represents momentum corresponding to the $P-$wave transition. The pion decay constant $ f_{\pi} $ = 132 MeV \cite{yan1992heavy} and the strong coupling constant $ a_1 = 0.612 $ as in Ref.\cite{pirjol1997predictions} obtained from quark model calculations.\\ 

\item {S$-$wave transitions}\\
 
\label{sec:3}

$S-$wave transitions of $\Lambda{_{c}^{+}} (1^{2}P_{\frac{1}{2}})$ into the isospin partners of $ {\Sigma{_{c}}} (1^{2}S_{\frac{1}{2}}) $ by an exchange of single pion are

\begin{equation}
\Gamma_{\Lambda{_{c}^{+}} \rightarrow \Sigma{{_{c}}{\pi}}} = \frac{b{_{1}^{2}}}{2\pi f{_{\pi}^{2}}} \frac{M{_{\Sigma{_{c}}}}}{M{_{\Lambda{_{c}^{+}}(1^{2}P_{\frac{1}{2}})}}} E{_{\pi}^2} p{_{\pi}}
\end{equation}

where $ p{_{\pi}} $ represent the $S-$wave transitions and when  single pion is at rest $ E{_{\pi}} \approx m{_{\pi}} $. The coupling constant $ b_1 = 0.572 $ and $b_2 = \sqrt{3}$ $ \cdot $ $b_1$ are taken from Ref.\cite{pirjol1997predictions}. The decay rates for the decay of isospin triplets $ \Sigma{_{c}}(1^{2}P_{\frac{1}{2}}) $ into $\Lambda{_{c}^{+}(1^{2}S_{\frac{1}{2}})} \pi $ are 

\begin{equation}
\Gamma_{\Sigma{_{c}} \rightarrow \Lambda{_{c}^+}\pi} = \frac{b{_{2}^{2}}}{2\pi f{_{\pi}^{2}}} \frac{M{_{\Lambda{_{c}^{+}}}}} {M{_{\Sigma{_{c}}}}} E{_{\pi}^2} p{_{\pi}}.
\end{equation}

\item {D$-$wave transitions}\\

\begin{table*}
\begin{center}
\caption{\label{tab:6}Strong one-pion decay rates (in MeV).}
\scalebox{0.85}{
%\begin{ruledtabular}
\begin{tabular}{ccccccccccccccccccc}
\hline\noalign{\smallskip}
 Decay mode & A & B & PDG\cite{partignani2016c}& \cite{cheng2015charmed}  & \cite{ivanov1999strong} &\cite{tawfiq1998charmed} & \cite{huang1995decays} & \cite{albertus2005study} &  \cite{pirjol1997predictions} & Others \\
\hline\noalign{\smallskip}
$P-$wave transitions\\
\hline\noalign{\smallskip}

$\Sigma{_{c}^{++}}(1^2S{_{\frac{1}{2}}})$ $\rightarrow $ $\Lambda{_{c}^{+}} \pi^{+}$ & 1.72 & 2.34 &  $ 1.89{_{-0.18}^{+0.09}} $ &   & 2.85 $ \pm $ 0.19 & 1.64 & 2.5 & 2.41 $ \pm $ 0.07  & 2.025 & $ 1.96{_{-0.14}^{+0.07}} $  \cite{Kawakami:2018olq}  \\

$\Sigma{_{c}^{+}}(1^2S{_{\frac{1}{2}}})$ $\rightarrow $ $\Lambda{_{c}^{+}} \pi^{0}$ & 1.60 & 2.59 & $<$ 4.6 & $2.3{_{-0.2}^{+0.1}}  $ & 3.63 $ \pm $ 0.27 & 1.70 &  3.2 & 2.79 $ \pm $ 0.08 & & $ 2.28{_{-0.17}^{+0.09}} $  \cite{Kawakami:2018olq} \\

$\Sigma{_{c}^{0}}(1^2S{_{\frac{1}{2}}})$ $\rightarrow $ $\Lambda{_{c}^{+}} \pi^{-}$ & 1.17 & 2.21 & $ 1.83{_{-0.19}^{+0.11}} $ & $1.9{_{-0.2}^{+0.1}}$  & 2.65 $ \pm $ 0.19 & 1.57 & 2.4 & 2.37 $ \pm $ 0.07  & 1.94 & $ 1.94{_{-0.14}^{+0.07}} $ \cite{Kawakami:2018olq} \\

\hline\noalign{\smallskip}

$\Sigma{_{c}^{*++}}(1^{4}S_{\frac{3}{2}})$ $\rightarrow $ $\Lambda{_{c}^{+}} \pi^{+}$ & 13.11 & 21.34 & $ 14.78{_{-0.40}^{+0.30}} $ & $14.5{_{-0.8}^{+0.5}}$ & 21.99 $ \pm $ 0.87 &  12.84 && 17.52 $ \pm $ 0.75  & 17.9 & $ 14.7{_{-1.1}^{+0.6}} $ \cite{Kawakami:2018olq} \\

$\Sigma{_{c}^{*+}}(1^{4}S_{\frac{3}{2}})$ $\rightarrow $ $\Lambda{_{c}^{+}} \pi^{0}$ & 14.28 & 12.83 & $<$ 17 & $15.2{_{-1.3}^{+0.6}}$ && &25  & 15.31 $ \pm $ 0.74  & & $ 15.3{_{-1.1}^{+0.6}} $ \cite{Kawakami:2018olq} \\

$\Sigma{_{c}^{*0}}(1^{4}S_{\frac{3}{2}})$ $\rightarrow $ $\Lambda{_{c}^{+}} \pi^{-}$ & 13.40 & 20.97 & $ 15.3{_{-0.5}^{+0.4}} $ & $14.7{_{-1.2}^{+0.6}}$  & 21.21 $ \pm $ 0.81 & 12.40 & & 16.90 $ \pm $ 0.72 & 13.0 & $ 14.7{_{-1.1}^{+0.6}} $ \cite{Kawakami:2018olq} \\

\hline\noalign{\smallskip}
$S-$wave transitions\\
\hline\noalign{\smallskip}

$\Lambda{_{c}^{+}(1^{2}P_{\frac{1}{2}})}$ $\rightarrow $ $\Sigma{_{c}^{++}} \pi^-$ & 3.92 & 5.54 &  & $0.72{_{-0.30}^{+0.43}}$  & 0.79 $ \pm $ 0.09 & 2.15 & $0.55{_{-0.55}^{+1.3}}$ & & & 0.64 \cite{zhu2000strong} \\

$\Lambda{_{c}^{+}(1^{2}P_{\frac{1}{2}})}$ $\rightarrow $ $\Sigma{_{c}^{0}} \pi^+$ & 4.45 & 5.63 & 2.6 $ \pm $ 0.6  & $0.77{_{-0.32}^{+0.46}}$ & 0.83 $ \pm $ 0.09 & 2.61 & $ 1.7 \pm 0.49 $ & & &  1.2 \cite{zhu2000strong}\\

$\Lambda{_{c}^{+}(1^{2}P_{\frac{1}{2}})}$ $\rightarrow $ $\Sigma{_{c}^{+}} \pi^0$ & 4.52 & 5.62 &  & $1.57{_{-0.65}^{+0.93}}$ & 0.98 $ \pm $ 0.12 & 1.73 & $ 0.89 \pm 0.86 $ & & & 0.84 \cite{zhu2000strong} \\

\hline\noalign{\smallskip}

$\Sigma{_{c}^{++}(1^{2}P_{\frac{1}{2}})}$ $\rightarrow $ $\Lambda{_{c}^{+}} \pi^+$ & 68.19 & 72.67 & $75{_{-17}^{+22}}$ & & & & & & & $ 75{_{-13-11}^{+18+12}}$ \cite{mizuk2005observation} \\

$\Sigma{_{c}^{+}(1^{2}P_{\frac{1}{2}})}$ $\rightarrow $ $\Lambda{_{c}^{+}} \pi^0$ & 62.92 & 64.54 & $62{_{-40}^{+60}}$ & & & & & & & $ 62{_{-23-38}^{+37+52}}$ \cite{mizuk2005observation} \\

$\Sigma{_{c}^{0}(1^{2}P_{\frac{1}{2}})}$ $\rightarrow $ $\Lambda{_{c}^{+}} \pi^-$ & 66.44 & 71.11 & $72{_{-15}^{+22}}$ & & & & & & &  $ 61{_{-13-13}^{+18+22}}$ \cite{mizuk2005observation} \\

\hline\noalign{\smallskip}
$D-$wave transitions\\
\hline\noalign{\smallskip}

$\Lambda{_{c}^{+}}{(1^{2}P_{\frac{3}{2}})}$ $\rightarrow $ $\Sigma{_{c}^{++}} \pi^{-}$ &  0.001  & 0.0012  &  & 0.029 & 0.076 $ \pm $ 0.009 & 2.15 & 0.013 & & & 0.011  \cite{zhu2000strong} \\

$\Lambda{_{c}^{+}}{(1^{2}P_{\frac{3}{2}})}$ $\rightarrow $ $\Sigma{_{c}^{0}} \pi^{+}$ &  0.011  &  0.0013  & $ < 0.97 $ & 0.029 & 0.080 $ \pm $ 0.009 & 2.61 & 0.013 & & & 0.011 \cite{zhu2000strong} \\

$\Lambda{_{c}^{+}}{(1^{2}P_{\frac{3}{2}})}$ $\rightarrow $ $\Sigma{_{c}^{+}} \pi^{0}$ &  0.033 &  0.0025  & & 0.041 & 0.095 $ \pm $ 0.012 & 1.73 & 0.013 & & & 0.011 \cite{zhu2000strong} \\

\hline\noalign{\smallskip}

$\Sigma{_{c}^{++}}{(1^{2}P_{\frac{3}{2}})}$ $\rightarrow $ $\Lambda{_{c}^{+}} \pi^{+}$ & 13.22 & 19.61 &&&&&& & $\sim 12$ \\ 

$\Sigma{_{c}^{++}}{(1^{4}P_{\frac{5}{2}})}$ $\rightarrow $ $\Lambda{_{c}^{+}} \pi^{+}$ & 10.68 & 15.91 &&&&&& & $\sim 12$ \\ 

$\Sigma{_{c}^{++}}{(1^{2}P_{\frac{3}{2}})}$ $\rightarrow $ $\Sigma{_{c}^{+}} \pi^{+}$ & 1.70 & 2.86 &\\ 

$\Sigma{_{c}^{++}}{(1^{2}P_{\frac{3}{2}})}$ $\rightarrow $ $\Sigma{_{c}^{*+}} \pi^{+}$ & 0.61 & 1.46 &\\ 

$\Sigma{_{c}^{++}}{(1^{2}P_{\frac{3}{2}})}$ $\rightarrow $ $\Sigma{_{c}^{*++}} \pi^{0}$ & 0.65 & 0.95 &\\ 

\hline\noalign{\smallskip}

\end{tabular}
}
\end{center}
\end{table*}

\label{sec:4}

The decay of $ \Lambda_{c}(1^{2}P_{\frac{3}{2}}) $ into the isospin partners of $\Sigma{_{c}(1^{2}P_{\frac{1}{2}}}) $ are consider as a $D-$wave transitions. For that the decay rates are

\begin{equation}
\Gamma_{\Lambda{_{c}^{+}}(1^{2}P_{\frac{3}{2}}) \rightarrow \Sigma{_{c}}\pi} = \frac{2b{_{3}^{2}}}{9\pi f{_{\pi}^{2}}} \frac{M{_{\Sigma{_{c}}}}}{M{_{\Lambda{_{c}^{+}}}}} p{_{\pi}^{5}}
\end{equation}

where $ p{_{\pi}^{5}} $ represents the $D-$wave transitions and the coupling constant $ b{_{3}} = $ 3.50 $\times$ $10^{-3}$ MeV$^{-1}$ Ref.\cite{pirjol1997predictions}. The $ \Sigma_c^{++}$ with $(1^{2}P_{\frac{3}{2}}) $ and $ \Sigma_c (1^{4}P_{\frac{5}{2}}) $; are expected to decay into $\Lambda{_{c}^{+}}(1^{2}S_{\frac{1}{2}}) \pi^{+}$ through $D-$wave couplings as 

\begin{equation}
\Gamma_{\Sigma{_{c}^{++}}\rightarrow \Lambda{_{c}^+}\pi^+} = \frac{4b{_{4}^{2}}}{15\pi f{_{\pi}^{2}}} \frac{M{_{\Lambda{_{c}^{+}}}}}{M{_{\Sigma{_{c}^{++}}}}} p{_{\pi}^{5}}
\end{equation}

here, the coupling constant $ b_{4} = 0.4 \times 10^{-3} $ MeV$^{-1}$ Ref.\cite{pirjol1997predictions}. According to the quark model relation, $ b_{5} = \sqrt{2 \cdot b_{4}} $. Using this, we obtained the decay rates for the decay of $ \Sigma_c^{++} (1^{2}P_{\frac{3}{2}}) $ into $\Sigma{_{c}^{+}}(1^{2}S_{\frac{1}{2}})\pi^{+}$, $\Sigma{_{c}^{*+}(1^{4}S_{\frac{3}{2}})} \pi^{+}$ and $\Sigma{_{c}^{*++}(1^{4}S_{\frac{3}{2}})} \pi^{0}$ are determine as \\

\begin{equation}
\Gamma_{\Sigma_{c}^{++} \rightarrow \Sigma{_{c}^+}\pi^+ / \Sigma{_{c}^{*}} \pi} = \frac{b{_{5}^{2}}}{10\pi f{_{\pi}^{2}}} \frac{M{_{\Sigma{_{c}^{+}/\Sigma{_{c}^{*}}}}}}{M{_{\Sigma{_{c}^{++}}}}} p{_{\pi}^{5}}.
\end{equation}

Summing up the decay rates of these three decay mode of $ {\Sigma{_{c}^{++}}( 1^{2}P_{\frac{3}{2}}}) $, it will be 2.97 MeV and 5.27 MeV for the set A and for set B, respectively; and the value of set A is nearer to $ \simeq 3.16 $ MeV of Ref.\cite{pirjol1997predictions}.  An obtained results for these three; S, P and D$-$wave transitions are listed in Table \ref{tab:6}.

\end{itemize}

\section{Discussion and Conclusion}

The electromagnetic radiative decays of singly charmed baryons by an exchange of massless photon are determined by using the parameters obtaining in the framework of hypercentral Constitute Quark Model (hCQM).\\

There are no experimental information available about the magnetic moments of singly charmed baryons. Our predictions of ground state magnetic moment of singly charmed baryons with $ J^P = {\frac{1}{2}}^+ $ and $ J^P = {\frac{3}{2}}^+ $ see Table \ref{tab:2} and Table \ref{tab:3} respectively; for the set A and set B are comparable to the results obtained from; bag model \cite{bernotas2013radiative}, effective quark mass scheme \cite{dhir2009magnetic}, non-relativistic quark model \cite{patel2008heavy}, chiral constitute quark model \cite{sharma2010spin} and relativistic quark model \cite{faessler2006magnetic}. For $ J^P = {\frac{3}{2}}^+ $, our results are smaller than the results based on QCD sum rule model \cite{aliev2009mass}. The recent paper of G. J. Wang et al. \cite{Wang:2018gpl} based on heavy chiral perturbation theory and K. U. Can et al. \cite{can2014electromagnetic}, H. Bahtiyar et al. \cite{bahtiyar2015omegacgamma,bahtiyar2017xicgamma} based on lattice QCD, their calculated magnetic moments for $ J^P = {\frac{1}{2}}^+ $ are lesser than our predictions.\\

The expression of electromagnetic radiative decay rate containing a term transition magnetic moment $ (\mu{_{{B{_{c}^{\prime}}} \rightarrow {B_c}}}) $ of the participating singly charmed baryons by which the decay is taking place. Our calculated transition magnetic moments and radiative decay rates smaller than other theoretical predictions. For $ \Xi{_{c}^{*+}} $, $ \Xi{_{c}^{*0}} $ and $ \Omega{_{c}^{*0}} $  our predictions are much smaller than others, and for $ \Sigma{_{c}^{*+}} $ the radiative decay rate is of the order of $ 10^{-1} $ to $ 10^{-5} $ keV, in our case it is $ 10^{-5} $ keV. Our results for the transition magnetic moment and radiative decay of $ \Sigma{_{c}^{+}} $, $ \Sigma{_{c}^{*++}} $, $ \Sigma{_{c}^{*+}} $ and $ \Sigma{_{c}^{*0}} $; are smaller but reasonable close to other theoretical predictions see Table \ref{tab:4} and Table \ref{tab:5}; respectively.\\

The strong $ P $-wave transitions of isospin partners $\Sigma_{c}(1^2S_\frac{1}{2})$ and  $ \Sigma{_{c}^{*}}(1^{4}S_{\frac{3}{2}}) $ are calculated, and are found to be in accordance with other model predictions and the experimental measurements. In our case, the ratio of $ \frac{\Gamma(\Sigma{_{c}^{*++}})}{\Gamma(\Sigma{_{c}^{++}})}$ is 7.62 for the set A  and 9.12 for the set B, and from the PDG \cite{partignani2016c} it is 7.82 consistent with set A. For the strong decay channel $ \Sigma{_{c}^{*}(1^{4}S_{\frac{3}{2}})} \rightarrow \Sigma{_{c}(1^2S{_{\frac{1}{2}}})} \pi $, the mass difference $ \Delta M{(m\Sigma{_{c}^{*}} - m\Sigma{_{c}})}$, is smaller than the mass of single pion. Therefore, there is no sufficient phase space for this respective decay. Such decay is kinematically forbidden.\\ 
 
For the $ S $-wave transitions of $\Lambda{_{c}^{+}} (1^{2}P_{\frac{1}{2}})$ that decay into isopartners of $ {\Sigma{_{c}}} (1^{2}S_{\frac{1}{2}}) $, are over estimated compare to others because here the mass of $\Lambda{_{c}^{+}} (1^{2}P_{\frac{1}{2}})$ is higher than the PDG  \cite{partignani2016c} value [Table \ref{tab:1}]. Also the decay of isotriplet $ \Sigma{_{c}}(1^{2}P_{\frac{1}{2}}) $ into $\Lambda{_{c}^{+}} \pi$, their decay widths are consistant with PDG \cite{partignani2016c} and Ref.\cite{mizuk2005observation}. The $ D $-wave transitions of $ \Sigma_c^{++}$ with $(1^{2}P_{\frac{3}{2}}) $ are decay into the various decay mode shown in Table \ref{tab:6}. The decay rates of $\Sigma{_{c}^{++}}(1^{4}P_{\frac{5}{2}})$ decay into $\Lambda{_{c}^{+}} (1^{2}P_{\frac{1}{2}}) \pi^{+}$ is also determined. Experimentally, these both the states are not confirmed yet and only few theoretical results are available. Whereas, the $ D $-wave transitions of $ \Lambda{_{c}^{+}}(1^{2}P_{\frac{3}{2}}) $ into the isospin partners of $ \Sigma{_{c}}(1^{2}P_{\frac{1}{2}}) $ are kinematically barely allowed having an extremely small width and this study will be useful for the experimental determination of their decay widths $ < $ 0.97 \cite{partignani2016c}.\\       

\noindent From these calculations we noted that the decay of $\Sigma{_{c}^{+}}(1^2S{_{\frac{1}{2}}})$ and $\Sigma{_{c}^{*+}}(1^{4}S_{\frac{3}{2}})$ into $\Lambda{_{c}^{+}}(1^{2}S_{\frac{1}{2}})$ are common into both strong and radiative decay. So we are interested to calculate their total decay width and the branching fractions.\\

The total decay rate is simply the sum of the decay rates of all individual decay. The branching fraction for particular decay mode is the ratio of the decay rate of particular decay rate to the relatively total decay rate. For example, the total decay width of $ \Sigma{_{c}^{+}}(1^2S{_{\frac{1}{2}}}) $

\begin{equation}
\Gamma_{tot(\Sigma{_{c}^{+}})}=\Gamma_{\Sigma{_{c}^{+}}(1^2S{_{\frac{1}{2}}})\rightarrow \Lambda{_{c}^{+}} \pi^{0}}  +  \Gamma_{\Sigma{_{c}^{+}}(1^2S{_{\frac{1}{2}}})\rightarrow \Lambda{_{c}^{+}} \gamma}
\end{equation}  

are $ \sim 1.66 $ MeV and $ \sim 2.66 $ MeV for the set A and set B, respectively; and the branching fraction of $ \Sigma{_{c}^{+}}(1^2S{_{\frac{1}{2}}})$ for the strong decay \\

\begin{equation}
{\cal{B}}_{\Sigma{_{c}^{+}}(1^2S{_{\frac{1}{2}}})\rightarrow \Lambda{_{c}^{+}} \pi^{0}}=\frac{\Gamma_{\Sigma{_{c}^{+}}(1^2S{_{\frac{1}{2}}})\rightarrow \Lambda{_{c}^{+}} \pi^{0}}}{\Gamma_{tot(\Sigma{_{c}^{+}})}}
\end{equation} 

are $ \sim 96.49$ $\% $ and $ \sim 97.49$ $\% $ for the set A and set B respectively. Similarly, the branching fraction of $ \Sigma{_{c}^{+}} (1^2S{_{\frac{1}{2}}}) $ for the radiative decay \\

\begin{equation}
{\cal{B}}_{\Sigma{_{c}^{+}}(1^2S{_{\frac{1}{2}}})\rightarrow \Lambda{_{c}^{+}} \gamma}=\frac{\Gamma_{\Sigma{_{c}^{+}}(1^2S{_{\frac{1}{2}}})\rightarrow \Lambda{_{c}^{+}} \gamma}}{\Gamma_{tot(\Sigma{_{c}^{+}})}}
\end{equation} 

are $ \sim 3.50$ $\% $ and $ \sim 2.50$ $\% $ for the set A and set B respectively.\\

In the same manner we determine the total decay rate of $\Sigma{_{c}^{*+}}(1^{4}S_{\frac{3}{2}})$, and they are $ \sim 14.42 $ MeV and $ \sim 12.96 $ MeV for the set A and set B respectively. For their strong decay, the branching fractions are $ \sim 99.00$ $\% $ and $ \sim 98.96$ $\% $ for the det A and set B respectively; and for the radiative decay they are $ \sim 1.00 $ $\% $ and $ \sim 1.04 $ $\% $ for the set A and set B respectively.\\

So we conclude that such a singly charmed baryons, $\Sigma{_{c}^{+}}(1^2S{_{\frac{1}{2}}})$ and $\Sigma{_{c}^{*+}}(1^{4}S_{\frac{3}{2}})$ are purely decay through strong interaction and is consistent with a PDG \cite{partignani2016c} value, $ \sim 100 $ $\% $.  We see the contribution of radiative decay is small to their total decay. Therefore, our results are accordance with present theoretical and the experimental status of singly charmed baryons; a the strong decays are dominant over the electromagnetic radiative decays. We hope that the future experiment like $ \bar{P}ANDA $ will be an unique position for providing the contribution of radiative decay for charm sector.\\

\noindent For the success of a particular model, it is required not only to produced the mass spectra but also decay property of these baryons. The masses obtained from hypercentral Constitute Quark Model (hCQM) are used to calculate the radiative and the strong decay widths. Such a calculated widths are reasonably close to the other model predictions and experimental observations (where available). This model has been successful in determining these properties, thus, we would like to use this scheme to calculate the decay rates of singly bottom baryons.

\end{document}